# Ti$_{n+1}$C$_n$ MXene with fully saturated and thermally stable Cl terminations

J. Lu[1], I. Persson[1], H. Lind[1], M. Li[2], Y. Li[2], K. Chen[2], J. Zhou[1,2], S. Du[2], Z. Chai[2], Z. Huang[2], L. Hultman[1], J. Rosen[1], P. Eklund[1], Q. Huang[2], and P.O.Å. Persson[1]

[1] Thin Film Physics Division, Department of Physics, Chemistry and Biology (IFM), Linköping University, SE-581 83 Linköping, Sweden

[2] Engineering Laboratory of Advanced Energy Materials (FiNE Lab.), Ningbo Institute of Industrial Technology, Chinese Academy of Sciences, Ningbo, Zhejiang, 315201, China

Corresponding author: per.persson@liu.se

**Abstract**

MXenes are a rapidly growing family of 2D materials that exhibit a highly versatile structure and composition, allowing for significant tuning of the material properties. These properties are, however, ultimately limited by the surface terminations, which are typically a mixture of species, including F and O that are inherent to the MXene processing. Other and robust terminations are lacking. Here, we apply high-resolution scanning transmission electron microscopy (STEM), corresponding image simulations and first-principles calculations to investigate the surface terminations on MXenes synthesized from MAX phases through Lewis acidic melts. The results show that atomic Cl terminates the synthesized MXenes, with mere residual presence of other termination species. Furthermore, *in situ* STEM-electron energy loss spectroscopy (EELS) heating experiments show that the Cl terminations are stable up to 750 °C. Thus, we present an attractive new termination that widely expands the MXenes' functionalization space and enable new applications.

**Introduction**

MXenes constitute a rapidly growing addition [1,2] to the family of 2D materials with excellent properties and performance in terms of electrochemical charge storage,[3,4] electromagnetic interference shielding,[5] filtering,[6] and more recently also carbon capture[7], in addition to a range of other applications.[8,9] MXenes are predominantly obtained from the parent inherently nanolaminated $M_{n+1}AX_n$ (MAX) phases where M is a transition metal, A is a group A element (mostly group 13 and 14) and X is C and/or N.[10] By chemical etching of the atomically thin A element layers that interleave sheets of $M_{n+1}X_n$. these sheets are separated from each other and their surfaces are immediately functionalized by surface terminating species, $T_X$,[11,12] that typically originate from the etchant. Accordingly, $M_{n+1}X_nT_X$ most appropriately describe MXenes. From this formula, it is apparent that the MXene properties can be tuned through variations in structure, composition, and surface terminations. The structure is inherited from the parent MAX phase (hexagonal, space group $P6_3/mmc$), but compositional tuning display an extraordinary toolbox for property tuning. This can be achieved through MXenes based on single M and X elements, as well as alloys on both M and X.[2,13] In addition, there are reports on MXenes forming out-of-plane [14] and in-plane [15] double-M elemental ordering, as well as vacancy-ordered structures.[16,17]

Manipulation of the surface terminations constitute the final and arguably most important variable for property tuning, [18] yet it is the experimentally least explored. Currently, the MXene preparation dictates that $T_X$ is inherent to the etchant and thus predominantly a combination of O and F, whose ratio can be tailored [19], where also OH has been considered as a minor [20] or even negligible contribution.[21] Despite several theoretical investigations considering single,[18,22] mixed,[23,24] and non-inherent [25-27] termination species, particularly the non-inherent terminations have remained relatively unexplored experimentally. Persson et al. explored a route for removing the inherent surface terminations through a combination of heating and $H_2$ etching, while also re-terminating the MXene surface from the gas phase with non-inherent species (e.g. $CO_2$).[7] However, to further expand the already vast property space for MXenes, non-traditional synthesis routes as well as new terminations must be considered.

In the present communication, we report on MXenes that are exclusively Cl terminated by synthesis from MAX phases through Lewis acidic melts. These results are significant, not only because they demonstrate Cl as an exclusive surface-terminating species for MXene, but also because the Cl terminations are highly stable, and thus offers a novel avenue for tailoring the properties of MXenes.

**Experiments**

MXene powder samples were synthesized by mixing homemade MAX phase powders, containing both $Ti_2AlC$ and $Ti_3AlC_2$, with $ZnCl_2$ and heated at 550 °C for 5 h. The resulting bulk product was crushed into powder and cleansed with aqueous HCl at ambient temperature. Then, the reacted material was obtained and purified with deionized water. The final powder sample was analyzed by scanning electron microscopy (SEM) to obtain the microstructural and chemical composition of the obtained material. MXenes and residual phases were identified by X-ray diffraction (XRD, D8 Advance, Bruker AXS, Germany) using Cu Kα radiation. For detailed information on the preparation process, please see ref. [28].

The atomic structural and chemical analysis were carried out by high-resolution high angle annular dark field scanning transmission electron microscopy (HAADF-STEM) imaging and lattice resolved energy dispersive X-ray (EDX) spectroscopy using the Linköping double Cs corrected FEI Titan[3] 60-300 microscope, operated at 300 kV and equipped with a high sensitivity Super-X EDX detector. Corresponding HAADF-STEM images were simulated using the Dr. Probe software for imaging conditions with high accuracy matching the experimental setup. Supercells of $Ti_2CCl_2$ and $Ti_3C_2Cl_2$ were constructed employing CrystalMaker with lattice parameters obtained from DFT calculations reported in the present paper.[29]

*In situ* heating of the samples was carried out using a furnace type heating holder (Gatan Model 652). EELS spectra were continuously acquired during heating using a Gatan Quantum ERS GIF. Resulting spectra were processed by background subtraction and plural scattering deconvolution routines embedded in the Digital Micrograph software.

Our calculations were carried out on primitive cells of Cl-terminated $Ti_2C$ and $Ti_3C_2$. A total supercell height of 40 Å was used to avoid self-interaction between MXene sheets across the periodic boundaries. We carried out first-principle calculations of total energies, density of states and band structures using Density Functional Theory (DFT) with plane wave methods as implemented in the Vienna *ab initio* Simulation Package (VASP) software.[30,31] The exchange-correlation effects were treated within the generalized gradient approximation (GGA) of Perdew-Burke-Ernzerhof (PBE).[32] The cutoff energy for the plane-wave expansion was set to 400 eV. A 25x25x1 mesh was used during structural relaxation of 1x1 $Ti_3C_2$ and $Ti_2C$. Such relaxations were carried out on multiple lattice parameters (in-plane a) to determine the equilibrium size, the results of which were fitted to a Morse-type equation of state.[33] Once an equilibrium lattice parameter was determined, accurate density of states and band structure were calculated using larger meshes of 29x29x1 for DOS, while band structure was calculated at 120 points between each pair of high symmetry points. $Cl_2$ molecules were calculated as well, by placing the molecule in a box 20 Å in size, we determined its bond length to 1.9928, which corresponds well to the experimentally expected value of 1.99.

**Results**

Fig. 1(a) shows the atomically resolved structure of a thin edge from a larger multilayer particle. In the thinnest areas, the structure exhibits the characteristic honeycomb appearance of a single sheet $M_2XT_x$ MXene, where bright dots correspond to the M element (Ti). The X element (carbon) is positioned in the core of the honeycomb, but is barely visible due to the mass contrast experimental conditions (intensity~$Z^2$). Electron diffraction patterns obtained from the same particle further confirm a hexagonal structure (see figure S1). An overview image of a multilayer particle with corresponding EDX elemental maps for Ti and Cl are shown in fig 1 b-d) respectively. Residual O was found in the EDX spectra but with no contribution from other elements. The EDX maps (see figure S2 for spectra) demonstrate that the particles are composed of these two elements, which additionally indicate that Cl is present as a surface termination. The location of the Cl termination can be inferred from the high resolution image in fig. 1a). Given that little contrast originates from the plan view observed honeycomb core, corresponding to the hcp site, then the fcc site is indirectly identified as the preferred site for Cl terminations. A STEM image simulation of a Cl-terminated $Ti_2C$ single sheet, with the terminations residing on the fcc site, is shown in the inset in fig. 1a (middle right). The simulation shows that the atomic columns (bright dots corresponding to Ti+Cl atoms in plan view) exhibit a homogeneous contrast, which matches the STEM image qualitatively.

Additional cross-sectional structural characterization was performed from similar particles as shown in fig. 2. The figure shows the layered characteristics of terminated $Ti_2C$ (fig. 2a,c) and

Ti$_3$C$_2$ (fig. 2 b,d) MXene from [11-20] and [1-100] orientations, respectively. The STEM images show that the sample consists of neatly stacked MXene sheets, each exhbiting bright atomic columns in the middle of the sheet and two darker surface layers. According to the inserted lattice resolved EDX elemental maps and by overlaying Ti (red) and Cl (green) it is clear that the darker surface termiantions consist of Cl, as seen from both the M$_2$X and M$_3$X$_2$ structures. Minor amounts of O was also observed during cross-sectional mapping. According to the EDX measurement, the Ti:Cl ratio is very close to 2:2 and 3:2, which directly corresponds to the observed number of layers and indicates an exclusively Cl-terminated Ti$_3$C$_2$ and Ti$_2$C MXene.

It is worth noting that the surface terminations are highly ordered, e.g. seemingly accepting equillibrium lattice positions. In previous observations of O- and F-terminated MXene [21], it was found that the terminations exhibit a significant degree of disorder on material prepared at ambient conditions. Here, the material was prepared during a high-temperature process, using temperatures equivalent to previous *in situ* observations of ordering among terminations,[21] which explains the highly ordered appearance.

The high degree of order among the surface terminations enables exact determination of the preferred site on the MXene surface. Combining the complementary information available from the [11-20] and [1-100] orientations, it is clear that Cl preferentially occupies the fcc site, in agreement with theoretical predictions.[34] This is further confirmed by the HAADF-STEM image simulations with Cl on the fcc site, which are overlaid as insets in figs 2 a-d) where the simulated intensities qualitatively matches the experimentally observed intensity difference between Ti and Cl layers.

However, from the STEM images above, it is also clear that the layer of Cl terminations is significantly separated from the nearest Ti layer, approximately 1.71 Å as measured for $Ti_2C$ from fig 2,c) and 1.84 Å for $Ti_3C_2$ from fig 2,b). This translates to a Ti-Cl bond length of 2.45 and 2.55 Å for an *a* lattice parameter of 3.05 Å in $Ti_2C$ and $Ti_3C_2$ respectively, which is typically much more than for other terminations. Previous experiments yield a Ti-O/F bond length of 2.11 Å for pristine $Ti_3C_2$ (translating to a Ti-O/F layer separation of 1.23 Å). [35]

Finally, the individual Cl-terminated MXene sheets are mutually translated along the basal plane compared to, e.g., O-terminated MXenes, where the fcc sites of two adjacent layers are located directly on top of each other, see fig 2b.

DFT calculations were performed for Cl-terminated $Ti_2C$ and $Ti_3C_2$, and it was found that Cl terminations prefer fcc sites over hcp or mixed for both structures in agreement with the experimental findings and previous calculations (see figure S3 for further information). Moreover, and in excellent agreement with the STEM images, the interplanar distance between the Cl terminations and the first Ti layer is exceptional and corresponds to 1.67 and 1.70 Å for $Ti_2C$ and $Ti_3C_2$, which corresponds to a calculated Ti-Cl bond length of 2.51 Å for both $Ti_2C$ and $Ti_3C_2$. In comparison, the Ti-C planes are separated by 1.09 and 1.03 Å ($Ti_2C$ and $Ti_3C_2$-nearest surface). See table S1 for calculated spacings.

Figure 3 shows the electronic structure, including band structure and projected density of states, determined using the fully relaxed structures of the MXene, with Cl terminations in the fcc sites. For both $Ti_2C$ and $Ti_3C_2$, the DOS and band structures are very similar, both showing well defined metallicity with the Fermi level dominated by Ti d-orbitals. In addition, there are strong indicators of hybridization between Ti d-, and C and Cl p-orbitals in the energy range between -2 and -7 eV. The difference between Ti$_2$C and Ti$_3$C$_2$ is mainly that there are more Ti and C bands in the Ti$_3$C$_2$ case. There is also a small gap in the band structure around -2 eV for Ti$_2$C, whereas in Ti$_3$C$_2$ that gap is closed.

Finally, the formation energy, $E_{form}$, of bonded Cl atoms on the surface was determined, as defined by the following formula:

$$E_{form} = E(\text{Ti}_{n+1}\text{C}_n\text{T}_2) - E(\text{Ti}_{n+1}\text{C}_n) - E(\text{T}_2)$$

Here, $E(\text{Ti}_{n+1}\text{C}_n\text{T}_2)$ refers to energy of the MXene surface with terminations T, while $E(\text{Ti}_{n+1}\text{C}_n)$ is the energy of the non-terminated surface, and $E(\text{T}_2)$ is the energy of a free molecule of the terminations (here Cl$_2$). According to this, the formation energy becomes -6.901 eV/formula unit (f.u). for Ti$_2$CCl$_2$, while the formation energy for Ti$_3$C$_2$Cl$_2$ is -6.694 eV/ f.u. This can be compared to the formation energy of O terminated Ti$_3$C$_2$O$_2$ and N terminated Ti$_3$C$_2$N$_2$, which are -9.656 and -2.596 eV/f.u., respectively, as shown in our previous article.[21] While not as stable as O, the Cl terminations can be expected to bond strongly to the MXene surfaces.

To experimentally investigate the calculated high stability of Cl terminations, the Ti$_3$C$_2$Cl$_2$ and the Ti$_2$CCl$_2$ powders were heated *in situ* up to 800 °C in the (S)TEM. The structural and elemental changes during heating were followed by (S)TEM imaging and electron energy loss spectroscopy (EELS). Fig. 4 shows the background-subtracted and plural-scattering-deconvolved EEL spectra from a Ti$_3$C$_2$Cl$_2$ particle as a variation of temperature. The spectra show the Cl-L$_{3,2}$ and Ti-L$_{3,2}$ edges at ~200 and ~450 eV energy loss, respectively, and are normalized against the Ti-L edge since the number of M layers does not change during heating. The EELS measurements additionally detected both carbon and oxygen, where additional carbon appears with temperature, presumably through beam induced contamination during the measurements and where the O is present (~5% ) from the initial measurement and remains stable. The results show that the $Ti_3C_2Cl_2$ is stable at temperatures below ~750 °C. Above this temperature, the MXene is gradually decomposing together with an apparent loss of Cl terminations. The loss is stronger at the thin edges of the MXene multilayer particle compared with the "bulk" of the particle, which is presumably due to a slow Cl diffusion-desorption process. After heating at 800 °C, only residual Cl remains in the multilayer particle. The loss of surface terminations from MXene during heating has been observed also for F, which desorbs at similar temperatures as Cl, but not for O which remains stable on the surface.[21] The high stability calculated for Cl terminations is therefore in agreement with the experimental findings. Note that the fine structure of the $Ti-L_{3,2}$ edge becomes smeared at the highest temperature which is presumably due to an increasing disorder among the remaining surface terminations and in the microstructure. The microstructure of the MXene sheets is indeed subject to degradation. While the laminated structure is preserved, an increasing amount of black spots appear, particularly apparent in the thinnest regions (see figure S4). These spots are interpreted as nanoscale voids, presumably formed by a local mechanical deformation of the MXene structure as Cl gas forms from the desorbing terminations.

**Discussion and Conclusions**

Cl has been investigated as a terminating species on $Ti_3C_2$ and $Ti_2C$ MXene. From high resolution (S)TEM imaging and lattice resolved EDX mapping, Cl was found to occupy the fcc sites of the MXene surface. In contrast to most etched MXenes, the surface terminations assume a highly ordered appearance, which is attained through the preceding high temperature processing.

A mix of terminations Cl, O and F, has been reported previously, where the MXene was etched from MAX powders using aqueous LiF-HCl [36] or aqueous HCl,[37] enforcing the mix of terminations. These termination-mixed MXenes resulted in an improved electrochemical rate capability, which was explained by an observed increase in interlayer spacing, causing improved ion accessibilty. The present results provide further support for this explanation. The very large separation observed here between the layer of Cl terminations and the underlying Ti surface is suggested to cause local undulations in a mixed termination layer that increases the average interlayer spacing. Theoretical calculations have predicted exclusively Cl-terminated $Ti_2C$ MXene to exhibit a very large voltage window (3.5-4 V),[34] which pinpoints the importance of further investigating routes for tailoring the MXene surface terminations.

*In situ* TEM heating further shows that the Cl terminations are thermally stable up to 750 °C, above which point they increasingly desorb, presumably as Cl gas, while the MXene structure additionally degrades through the formation of local nanoscale voids. After heating at 800 °C only residual Cl remain on the surface.

The present results identify MXenes, which are exclusively terminated by Cl. These were formed by MXene synthesis using Lewis acidic melts to extract the A-layer at high temperatures. This method shows promise as a feasible way to synthesize new MXenes with single-component surface terminations. Consequently, the present communication extends the range of available surface chemistries on MXenes and facilitates tailoring of the amount of Cl terminations, from mixed (O+F+Cl and O+Cl) to exclusively Cl-terminated. Moreover, these findings stress that MXene terminations are not limited to F and O and that new and more terminations remain to be identified.

**Acknowledgements**

The authors acknowledge the Swedish Research Council for funding under Grants Nos. 2016-04412 and 2013-8020, the Swedish Government Strategic Research Area in Materials Science on Functional Materials at Linköping University (Faculty Grant SFO-Mat-LiU No. 2009 00971), and National Natural Science Foundation of China (Grant No. 21671195 and 91426304). The Knut and Alice Wallenberg Foundation is acknowledged for support of the electron microscopy laboratory in Linköping, Fellowship grants and a project grant (KAW 2015.0043). P.O.Å.P. and J.R., also acknowledge Swedish Foundation for Strategic Research (SSF) through project funding (EM16-0004) and the Research Infrastructure Fellow RIF 14-0074.

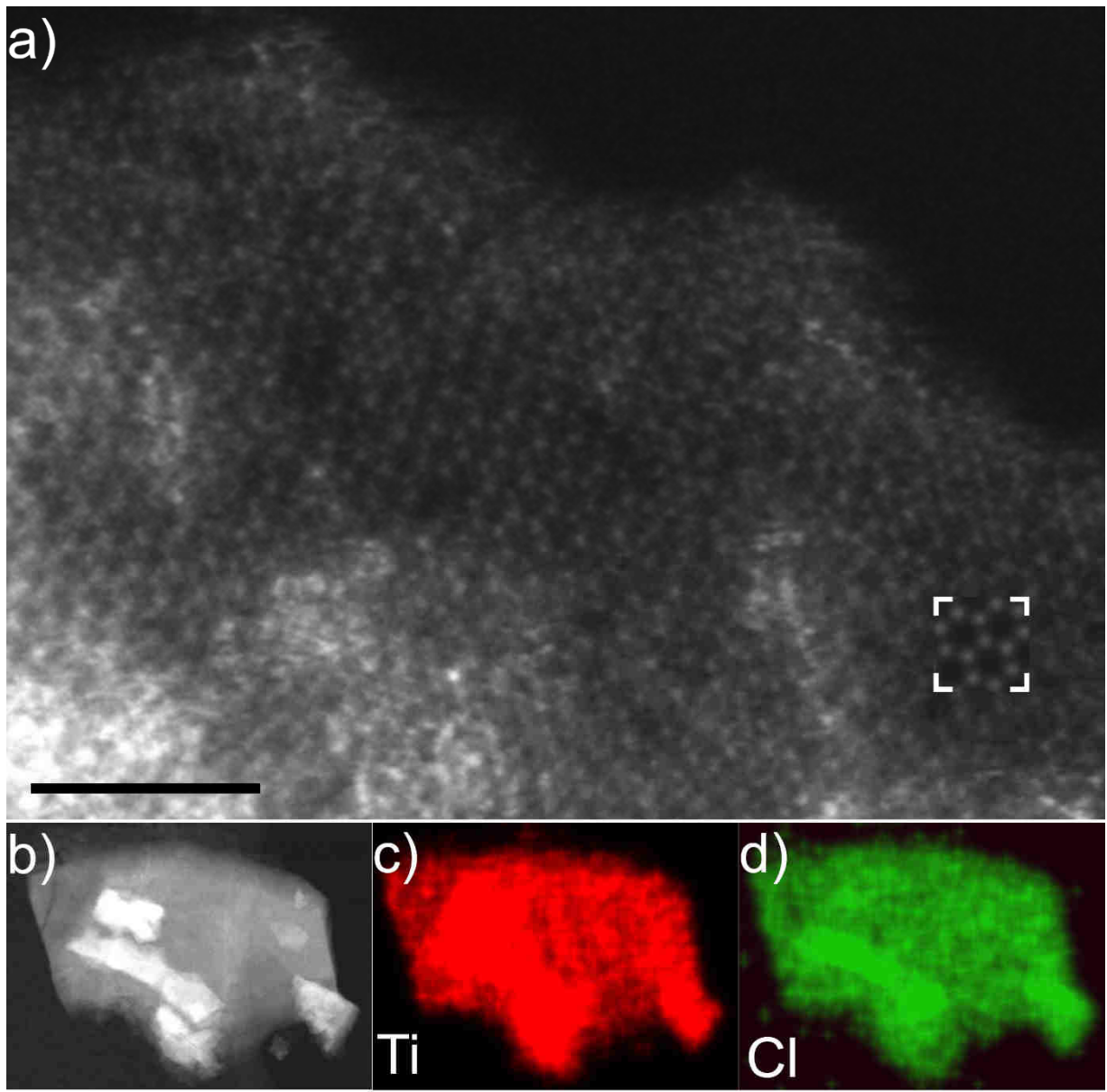

*Figure 1*. a) Atomically resolved plan-view HAADF-STEM image of a Ti$_2$CT$_x$ MXene particle and b) a low magnification overview with corresponding c) Ti and d) Cl elemental maps. Note the inset in a) showing a simulated plan-view stem image of Ti$_2$CCl$_2$ (middle right). The scalebar in a) correspond to 2 nm.

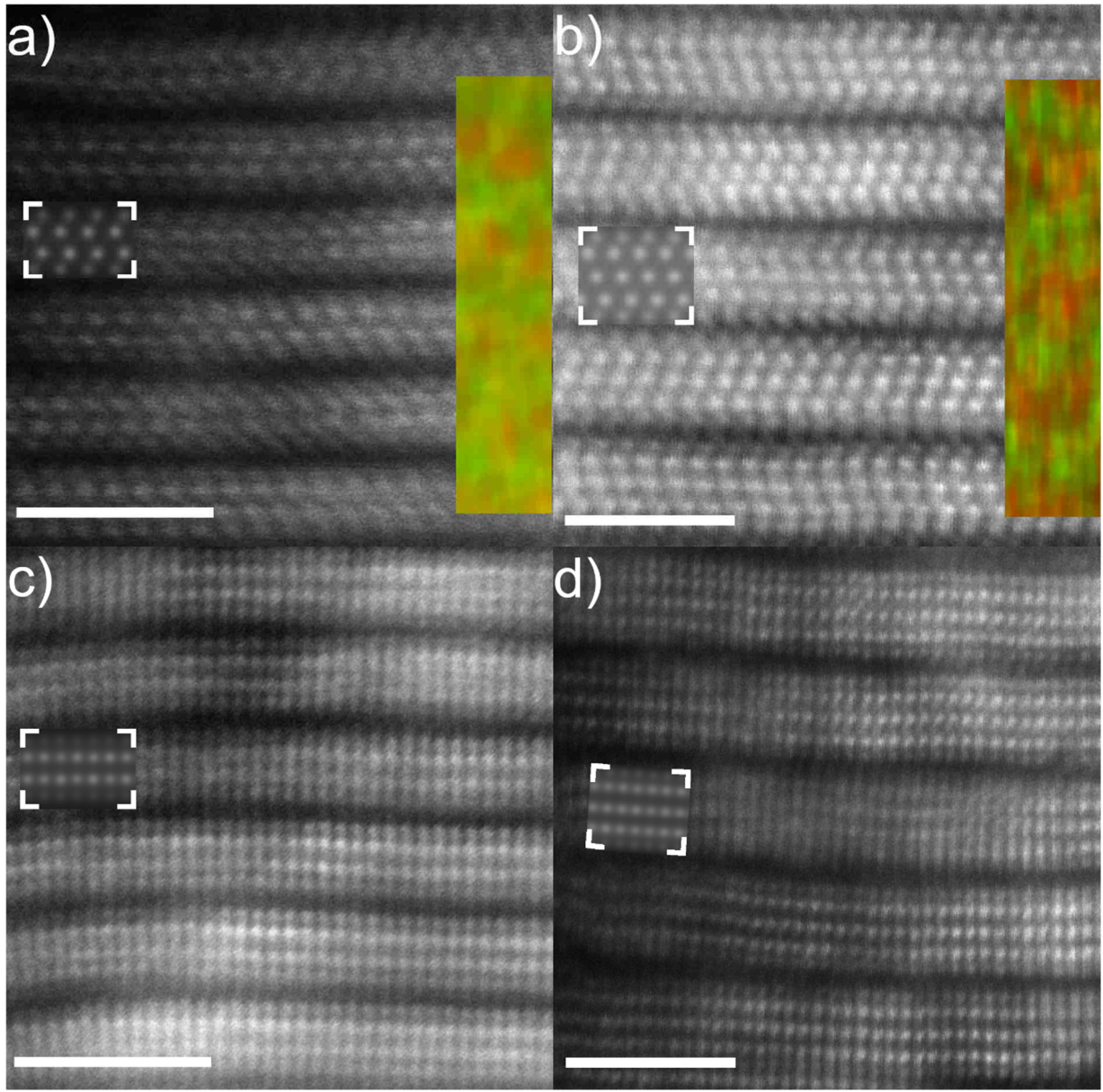

*Figure 2*. Atomically resolved cross sectional HAADF-STEM images for {11-20} oriented a) $Ti_2CCl_2$ and b) $Ti_3C_2Cl_2$, and for {1-100} oriented c) $Ti_2CCl_2$ and a) $Ti3C_2Cl_2$. The colored insets in a) and b) correspond to lattice resolved EDX elemetal maps showing Ti (red) and Cl (green). Note the insets showing simulated cross sectional STEM images of $Ti_2CCl_2$ and $Ti_3C_2Cl_2$ for each orientation. Scalebars correspond to 2 nm.

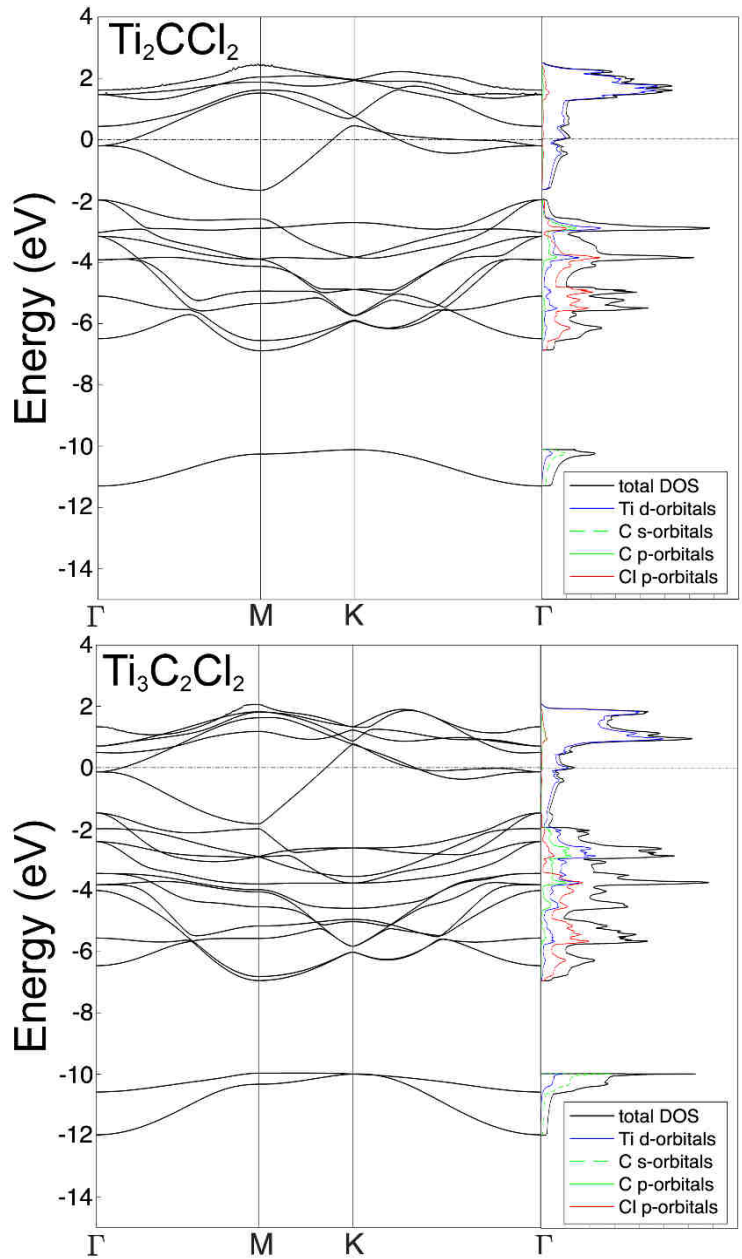

*Figure 3.* Band structure and electronic density of states for $Ti_2C$ (top) and $Ti_3C_2$ (bottom) with fcc coordinated Cl terminations.

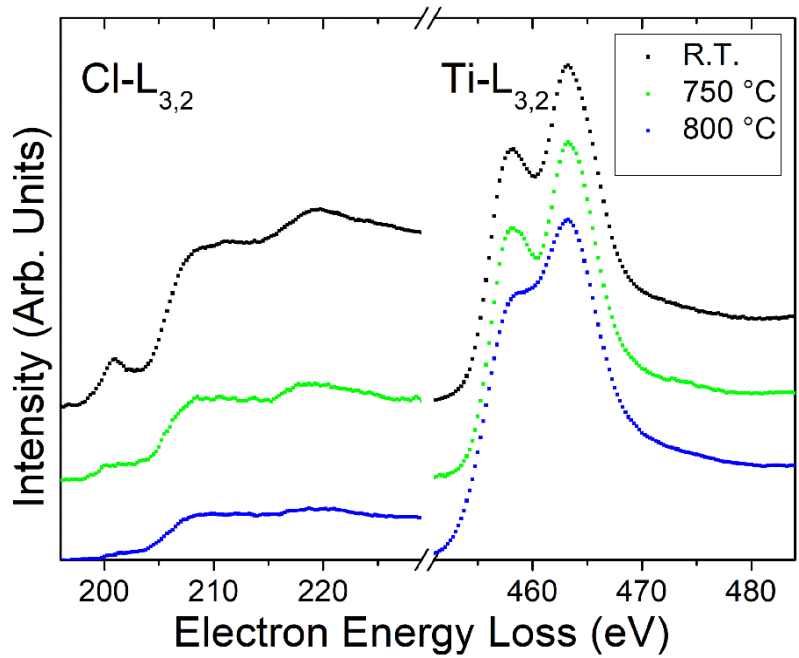

*Figure 4.* Electron energy loss spectroscopy from a thin edge of a $Ti_3C_2T_2$ multilayer particle, showing the background subtracted and multiple scattering deconvolved Cl- and Ti -$L_{3,2}$ edges at ambient conditions and after heating at 750 and 800 °C. The spectra are normailzed against the Ti –$L_2$ peak maximum and vertically separated for clarity.

**Ti$_{n+1}$C$_n$ MXene with fully saturated and thermally stable Cl terminations**

J. Lu[1], I. Persson[1], H. Lind[1], M. Li[2], Y. Li[2], K. Chen[2], J. Zhou[1,2], S. Du[2], Z. Chai[2], Z. Huang[2], L. Hultman[1], J. Rosen[1], P. Eklund[1], Q. Huang[2], and P. O.Å. Persson[1]

[1] Thin Film Physics Division, Department of Physics, Chemistry and Biology (IFM), Linköping University, SE-581 83 Linköping, Sweden

[2] Engineering Laboratory of Advanced Energy Materials (FiNE Lab.), Ningbo Institute of Industrial Technology, Chinese Academy of Sciences, Ningbo, Zhejiang, 315201, China

Corresponding author: per.persson@liu.se

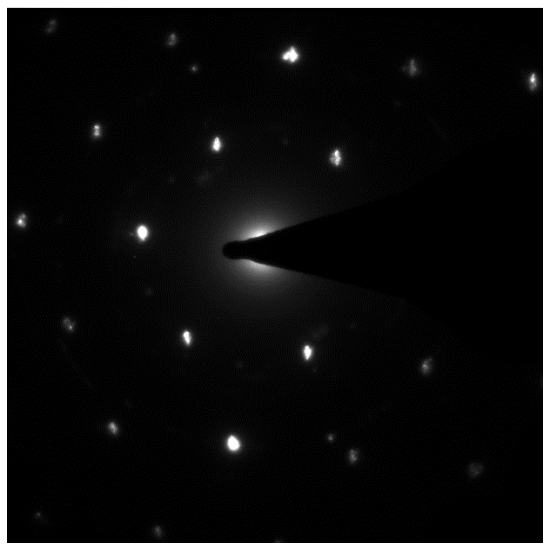

Figure S1. Electron diffraction pattern obtained from the multilayer particle shown in fig 1b

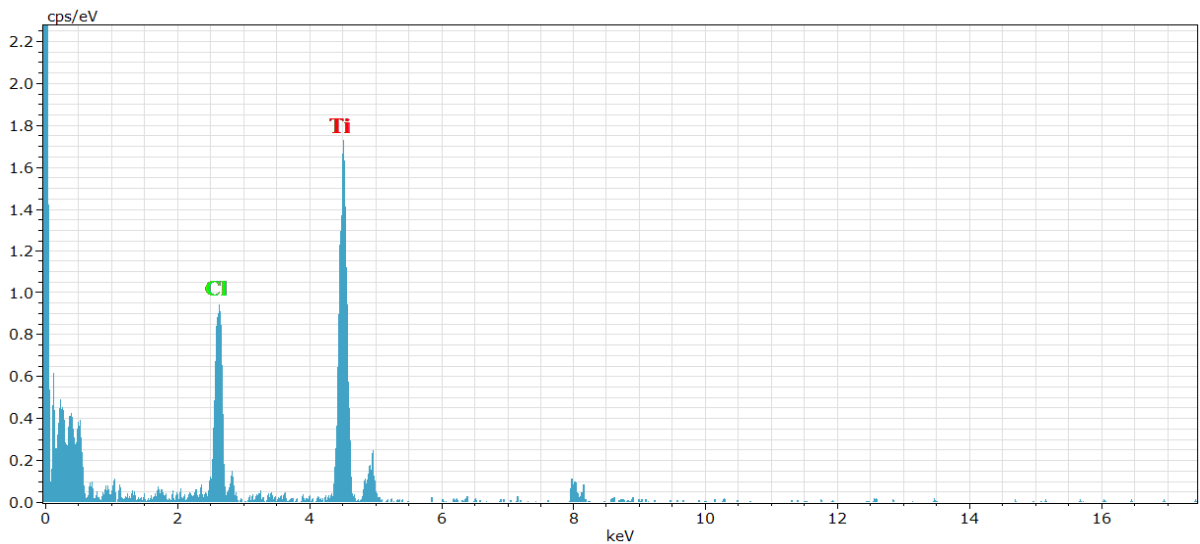

Figure S2. EDX spectrum obtained from the particle in fig 1b, showing evidence for the presence of Ti and Cl. The weak peak at 8.03 kV belongs to Cu Kα originating from the copper grid. Additionally, a small contribution from O is seen, originating from a $TiO_2$ particle attached to the surface of the MXene particle (elongated bright particle) in figure 1b.

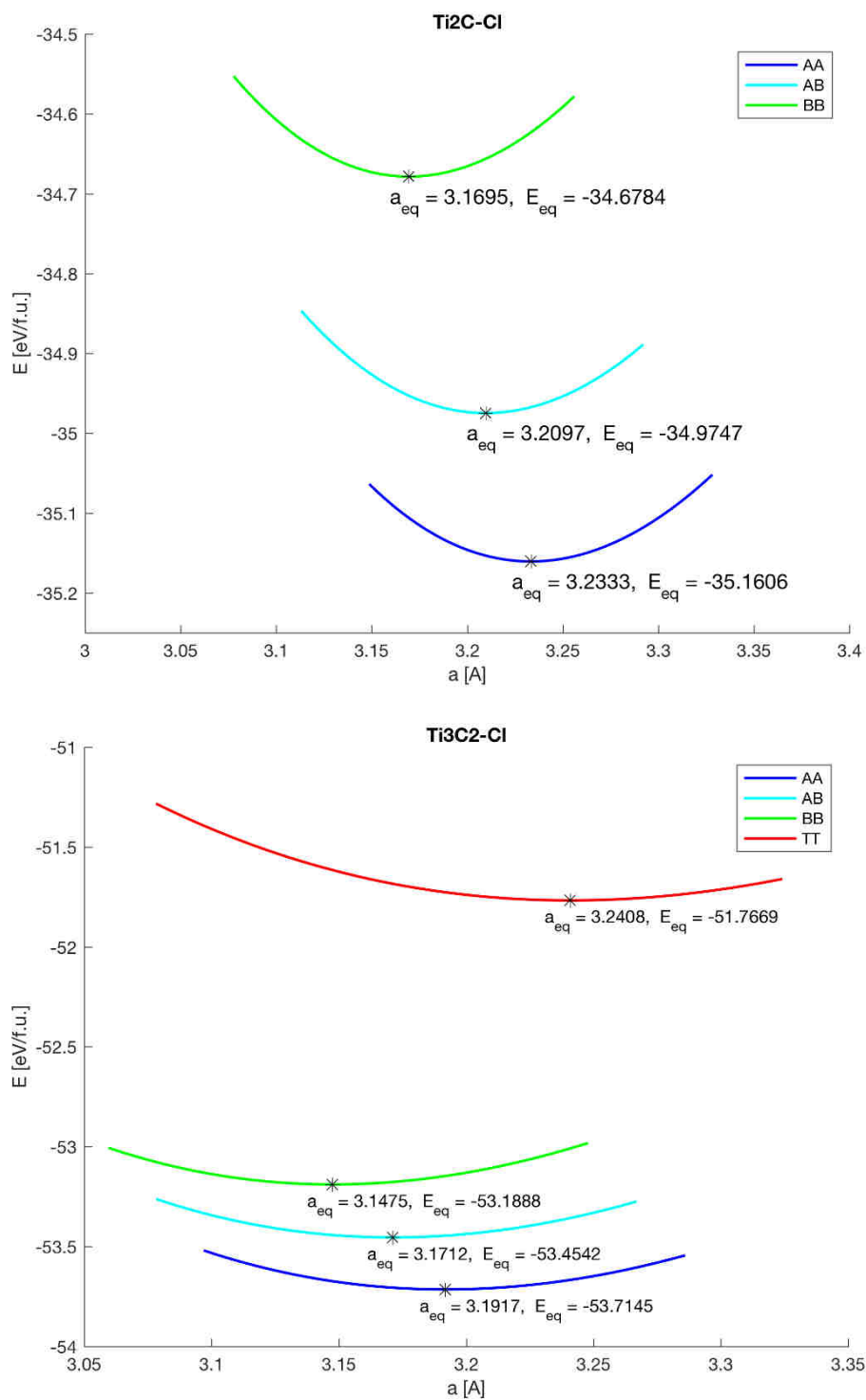

*Figure S3. Formation energy and optimum lattice parameters for Cl terminated $Ti_2C$ and $Ti_3C_2$ with the terminations residing on AA - fcc sites (top and bottom surfaces), AB – fcc (top surface) and hcp (bottom surface) sites as well as BB – hcp sites (top and bottom surface). For $Ti_3C_2$ the TT - atop configuration was also found to be stable.*

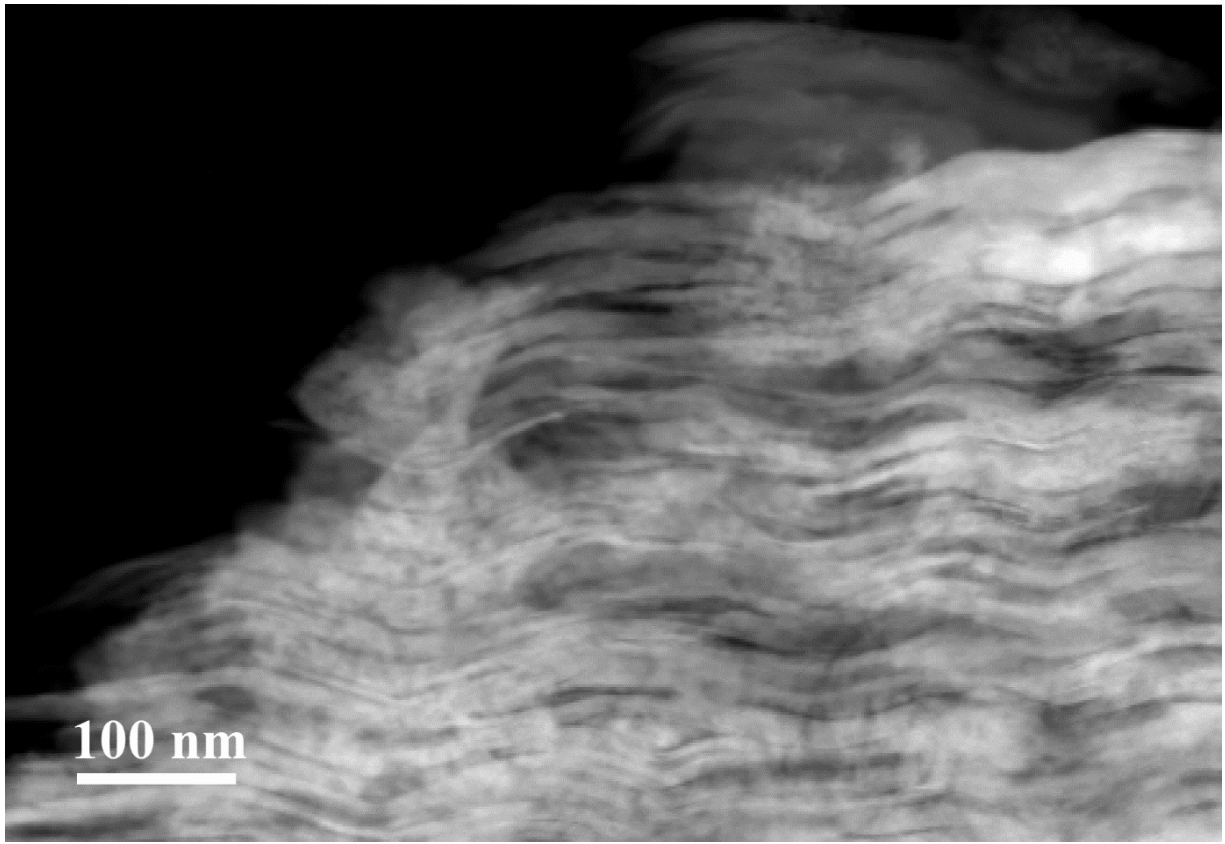

Figure S4. Low magnification STEM image of the as heated $Ti_3C_2Cl_X$ MXene, exhibiting local dark spots.

|  | Bond length | Planar spacing |
| --- | --- | --- |
| (Ti$_2$C) Ti-Cl | 2.5091 Å | 1.6766 Å |
| (Ti$_2$C) Ti-C | 2.1609 Å | 1.0885 Å |
| (Ti$_3$C$_2$) Ti1-Cl | 2.5086 Å | 1.7022 Å |
| (Ti$_3$C$_2$) Ti1-C | 2.1131 Å | 1.0341 Å |
| (Ti$_3$C$_2$) Ti2-C | 2.2320 Å | 1.2594 Å |

*Table S1. Calculated bond lengths and planar spacings for the fcc coordinated Cl terminated MXenes.*